\documentclass[a4paper,twoside,11pt]{article}
\newif\ifdraft \global\drafttrue
\def\production{\global\draftfalse}
\production
\usepackage[T1]{fontenc}
\usepackage[latin1]{inputenc}
\usepackage{a4wide}
\usepackage{times}
\usepackage{graphicx}
\usepackage{theorem, comment}
\usepackage[colorlinks=true,hyperindex=true]{hyperref}
\usepackage{url}
\usepackage{color}
\usepackage{fancyhdr}
\usepackage{latexsym}
\usepackage{amsmath}
\usepackage{bbm}
\usepackage{amssymb}
\usepackage{amsfonts}
\usepackage{enumerate}
\usepackage{tikz}
\usepackage{cancel}
\ifdraft
\usepackage{showkeys}
\fi
\newcounter{smallarabics}

\newcounter{smallroman}

\newcommand{\ben}{\begin{enumerate}[{\rm (1)}]}
\newcommand{\een}{\end{enumerate}}

\newtheorem{theoreme}{Theorem }[section]
\newtheorem{proposition}[theoreme]{Proposition}
\newtheorem{lemma}[theoreme]{Lemma}
\newtheorem{definition}[theoreme]{Definition}

\def\rr{{\mathbb R}}

\def\cc{{\mathbb C}}

\def\textsl{{}}

\def\Re{{\rm Re}\,}
\def\ch{\mathfrak{h}}

\def\c0inf{C_0^\infty}
\def\bep{\begin{proposition}}
\def\eep{\end{proposition}}

\def\cH{{\cal  H}}

\def\cR{{\cal R}}
\def\cA{{\cal A}}

\def\f{{\rm f}}

\def\i{{\rm i}}
\def\f{{\rm f}}
\newcommand{\beq}{\begin{equation}}
\newcommand{\eeq}{\end{equation}}
\newcommand{\bear}[1]{\begin{array}{#1}}
\newcommand{\ear}{\end{array}}

\def\sp{{\hat e}}

\newcommand{\e}{\mathrm{e}}
\renewcommand{\i}{\mathrm{i}}

\renewcommand{\d}{\mathrm{d}}

\def\cA{{\cal A}}

\def\bel{\begin{lemma}}
\def\eel{\end{lemma}}
\def\bet{\begin{theoreme}}
\def\eet{\end{theoreme}}
\def\bed{\begin{definition}}
\def\eed{\end{definition}}
\def\bar{\overline}

\def\12{\frac{1}{2}}

\def\e{{\rm e}}

\def\d{{\rm d}}

\def\cH{{\cal H}}

\def\sp{{\rm sp}}
\def\cS{{\cal S}}
\def\cR{{\cal R}}

\def\tr{{\rm tr}}

\def\P{\mathbb P}
\def\cO{{\cal O}}

\def\Jl0{J^{(l)}_{0}}
\def\Jr0{J^{(r)}_0}

\def\dl0{\delta_{-1}}
\def\dr0{\delta_{1}}

\def\chl0{\chi^{(l)}_0}
\def\chr0{\chi^{(r)}_0}
\def\ch0lr{\chi^{(l/r)}_0}
\def\de0{\delta_0}

\def\Jlr0{J^{(l/r)}_0}

\begin{document}
\def\today{}

\title{Full statistics of erasure processes:\\ Isothermal adiabatic theory\\ and a statistical Landauer principle}
\author{T. Benoist$^{1}$, M. Fraas$^2$, V. Jak\v{s}i\'c$^{3}$, C.-A. Pillet$^4$
\\ \\ 
$^1$ CNRS, Laboratoire de Physique Th\'eorique, IRSAMC,\\
Universit\'e de Toulouse, UPS, F-31062 Toulouse, France
\\ \\
$^2$Mathematisches Institut der Universit\"at M\"unchen\\ 
Theresienstr. 39, D-80333 M\"unchen, Germany
\\ \\
$^3$Department of Mathematics and Statistics\\ 
McGill University\\
805 Sherbrooke Street West \\
Montreal,  QC,  H3A 2K6, Canada
\\ \\
$^4$Universit\'e de Toulon, CNRS, CPT, UMR 7332, 83957 La Garde, France\\
Aix-Marseille Universit\'e, CNRS, CPT, UMR 7332, Case 907, 13288 Marseille, France\\
FRUMAM
}
\maketitle
\thispagestyle{empty}

\noindent{\bf Abstract.} We study driven finite quantum systems in contact with a thermal reservoir in the regime where
 the system changes slowly in comparison to the equilibration time.
  The associated 
isothermal adiabatic theorem allows us to control the full statistics of 
energy transfers in quasi-static processes. With this approach, we extend  
Landauer's Principle on the energetic cost of erasure processes to the level 
of the full statistics and elucidate the nature of the fluctuations breaking Landauer's bound.
\bigskip
\section{Introduction}
\label{SEC-Intro}

Statistical fluctuations of physical quantities around their mean values are
ubiquitous phenomena in microscopic systems driven out of thermal equilibrium. 
Obtaining the full statistics (i.e., the probability distribution) of 
physically relevant quantities is essential for a complete understanding of 
work extraction, heat exchanges, and other processes pertaining to these 
systems. The dominant theoretical and experimental focus in this respect is on 
classical~\cite{ECM,Cr, GC, Ja} and quantum~\cite{Ta,Ku}
{\sl universal fluctuation relations.} In our opinion, a task of equal importance is to 
derive from first principles and experimentally verify the Full Statistics of energy 
transfers in paradigmatic non-equilibrium processes.

In this note, we contribute to this task by investigating the Full
Statistics of the heat dissipated during an erasure process~\cite{DL} in the
adiabatic limit. While the expected value of this quantity is bounded below by the 
celebrated {\sl Landauer Principle,} we show that its Full Statistics possesses extreme outliers: albeit with a small probability, a large heat current breaking Landauer's
bound might be observed. The signature of this phenomena is a singularity of the
cumulant generating function of the dissipated heat. Our findings give an additional 
theoretical prediction to look for in experiments~\cite{SSS, PGA} investigating the Landauer
bound. The core tool in deriving our results, which is of theoretical interest
on its own, is a Full Statistics adiabatic theorem.

{\bf Thermodynamics and Information.} Thermodynamics and statistical mechanics are intimately
linked with information theory through  an intriguing world of infernal
creatures, thought experiments and analogies. In this world, Maxwell's demon is
effortlessly decreasing the Boltzmann entropy of an ideal gas~\cite{Wib}, and
the Szilard engine is converting the internal energy of a single heat bath into
work~\cite{Wia}. Both processes are in apparent violation of the second law of
thermodynamics. This fundamental law is, however, restored by considering the
Shannon entropy of the information acquired by the beings of this world during
these processes.

Landauer's thesis that {\sl information is physical}~\cite{La} provides a
guiding principle for exploring  the paradoxes of the aforementioned world. Together
with Bennett~\cite{Be} they argue that information is stored on physical devices
and hence its processing has to obey the laws of physics. A bit, an abstract binary
variable with values $0/1$, can be implemented by a charge
in a capacitor, or a colloidal particle trapped in a double-well. A two level
quantum system, called a qubit, can be physically realized by  the two energy levels in a
quantum dot or in a trapped atom. Irrespective of the realization, the laws
of mechanics and thermodynamics apply to these devices.

\begin{figure}
\label{FIG-erasure}
\begin{center}
\includegraphics[scale=0.5]{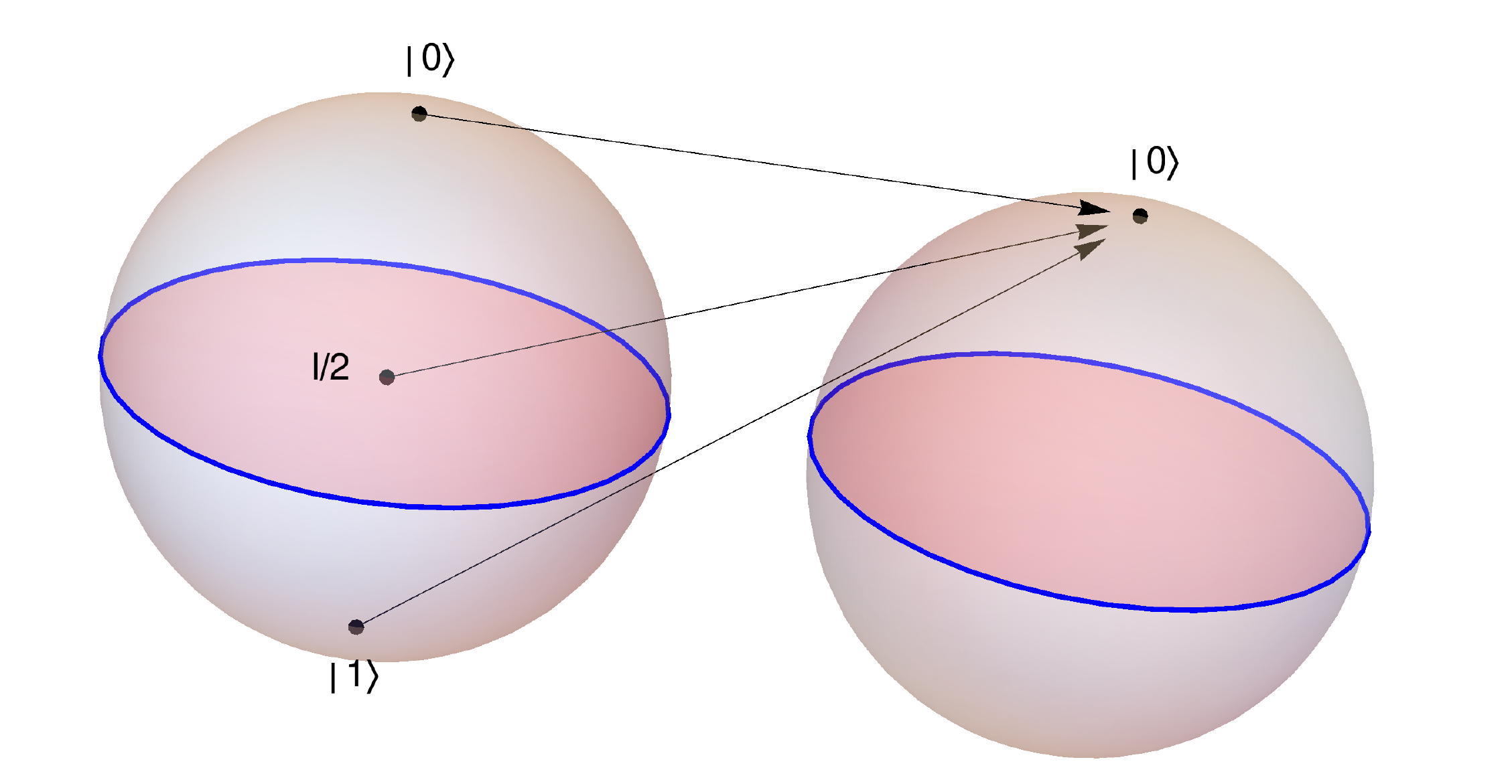}
\end{center}
\caption{An erasure process maps the Bloch ball, the phase space of a qubit,
into a single pure state, e.g., the point $\left|0\right\rangle$. A
measurement of the qubit after erasure would not provide any information
about the initial state. Since a general quantum operation transforms the Bloch ball into a
(possibly degenerate) ellipsoid centered at the image of $I/2$, a process is an erasure process if and only if it
maps the completely mixed state $I/2$ into a pure state.}
\end{figure}

Conservation of the phase space, in particular, implies that reversible
operations (such as  the gate  mapping $0\to1$ and vice versa)  can be produced with an arbitrary small energy cost, while any 
irreversible operation would dissipate a certain minimal amount of heat. A
paradigmatic example of the latter, invoked when erasing the memory of Maxwell's
demon \cite{Be}, is the erasure process (see Figure~\ref{FIG-erasure}).
In this process the entire phase space is mapped into a single point. The minimal amount of 
heat dissipated during this operation is described by the following principle, due to Landauer.

{\bf Landauer's Principle.} Specializing to quantum mechanics, we
consider the transformation of  an initial state $\rho_\i$ of a qudit\footnote{A quantum system described by a d-dimensional Hilbert space.} $\cS$
into a final state $\rho_\f$. If the initial/final entropies
$S_{\i/\f}=-\tr(\rho_{\i/\f}\log\rho_{\i/\f})$ are distinct, then the
transition $\rho_\i\to\rho_\f$ is irreversible: it can only be realized
by coupling the system $\cS$ to a reservoir $\cR$. The Landauer principle gives
a lower bound for the energetic cost of such a transformation in cases
when the reservoir is a thermal bath in equilibrium at a given temperature 
$T=1/k_B\beta$. The average heat $\langle\Delta Q\rangle$ dissipated in the 
reservoir by an arbitrary process which instigates the transition 
$\rho_\i\to\rho_\f$ is bounded from below as
\begin{equation}
\label{EQ-Landauer}
\beta  \langle \Delta Q \rangle \geq  \Delta S,
\end{equation}
where $\Delta S=S_\i-S_\f$ is the entropy difference. If $\rho_\i$ is the completely 
mixed state, and $\rho_\f$ a pure state, the above process is called {\sl erasure.} 
For system with $d$-dimensional Hilbert space, we then have $S_\i=\log d$, 
$S_\f=0$, and the Landauer bound takes the simple form
\beq
\label{EQ-erasure}
\beta\langle\Delta Q\rangle\geq \log d.
\eeq
The Landauer principle is a reformulation of the second law of quantum thermodynamics for qudits~\cite{RW, JP}. 
This can be immediately deduced from the entropy 
balance equation of the process
\beq
\label{EQ-EntBal}
\Delta S+\langle\sigma\rangle=\beta\langle\Delta Q \rangle.
\eeq
In one direction, the second law stipulates that the entropy
production $\langle \sigma \rangle$ is non-negative, implying Eq.~\eqref{EQ-Landauer},
and, in the opposite direction, Inequality~\eqref{EQ-Landauer} implies that
$\langle\sigma\rangle\geq 0$. A microscopic derivation of the Landauer
Principle was recently given in~\cite{RW} for finite dimensional reservoirs and
in~\cite{JP} for infinitely extended reservoirs. Both works use an appropriate,
rigorously justified version of the entropy balance equation~\eqref{EQ-EntBal}.\footnote{In a recent work \cite{HJPR}, 
the entropy balance equation has been used for study of Landauer's principle in repeated interaction systems.}
Landauer's Principle was also experimentally confirmed in several classical systems~\cite{SLKL,PBVN,Ra,TSU,BAP}; see also 
the recent reviews~\cite{LC, Pe}.

Processes involving only finite-volume reservoirs cannot saturate 
Inequality~\eqref{EQ-Landauer}. In fact, tighter lower bounds can be derived
in these cases: see~\cite{RW} and~\cite{GPaM}. In the thermodynamic limit, however,
equality is reached by some reversible quasi--static processes~\cite{JP}.
Such a process is realized by a slowly varying Hamiltonian 
$$
[0,T]\ni t\mapsto H_{\cS}(t/T)+H_{\cR}+\lambda(t/T)V
$$
along any trajectory in the parameter space such that $\lambda(0/1)= 0$ and 
\beq
\label{EQ-Hsys}
H_{\cS}(0/1)=-\beta^{-1}\log\rho_{\i/\f}+F_{\i/\f}I.
\eeq
Here, $H_\cR$ denotes the Hamiltonian of the reservoir and $V$ its coupling
to the system $\cS$, while  $F_{\i/\f}$ are  arbitrary constants.
In the adiabatic limit $T\to\infty$, the unitary evolution generated by the
corresponding Schr\"{o}dinger equation on the time interval $[0,T]$
transforms the initial state  $\rho_{\rm i}$ of the system  ${\cal S}$ to its final state $\rho_\f$, and the equality
holds in~\eqref{EQ-Landauer}. The quantity $\Delta F=F_\f-F_\i$ is endowed with the
meaning of a free energy difference.

{\bf Heat Full Statistics.} In this work we study the fine balance between heat
$\Delta Q$ and entropy $\Delta S$ in such quasi--static transitions, {\sl beyond 
the average value} $\langle\Delta Q\rangle$. To saturate Landauer's 
bound, we have to work with infinitely extended reservoirs and infinitely slow driving
forces, so the definition of $\Delta Q$ is  subtle.

The notion of Full Statistics (FS) was introduced in the study of quantum
transport~\cite{SS,LL1,LL2,LLY} (see also~\cite{ABGK,JOPP} for more mathematically 
oriented approaches) in order to characterize the charge fluctuations in mesoscopic 
conductors in terms of higher cumulants of their statistical distribution.
The later extension of FS to a more general setting, including energy transfers, 
led to the formulation of fluctuation relations in quantum physics~\cite{Ku,Ta}. In 
this approach, energy variations are not associated to a single observable~\cite{TLH}
but to a two-time measurement protocol.

Following the works of Kurchan and Tasaki~\cite{Ku,Ta} we identify the FS of the
dissipated heat $\Delta Q$ with that of the variation in the reservoir energy
during the process. This variation is {\sl defined} as the difference between the 
outcomes of two energy measurements: one at the initial time $0$ and another one at the 
final time $T$. The FS of $\Delta Q$ is the probability distribution (pdf) of the
resulting {\sl classical} random variable. The detailed derivation of 
this FS is given in Section~\ref{SEC-Setup}. 

We want to emphasize that
the extended reservoir has infinite energy and a continuum of modes. Consequently, to
obtain the FS of $\Delta Q$ one has to start with finite reservoirs and perform
the thermodynamic limit of the measurement protocol. In this limit, and for
generic processes, the random variable $\Delta Q$ acquires a continuous range.

Our main result is an explicit formula for the probability
distribution of $\Delta Q$ in the above quasi--static processes which saturate the
erasure Landauer bound~\eqref{EQ-erasure}. We show that for the completely mixed initial 
state $\rho_\i=I/d$ and a  strictly positive final state $\rho_\f$ the 
cumulant generating function of the dissipated heat is given by
\begin{equation}
\label{mainresult}
\log\langle\e^{-\alpha\Delta Q}\rangle
=-\frac\alpha\beta\log d+\log\tr\left(\rho_\f^{1-\frac\alpha\beta}\right).  
\end{equation}
Assuming for simplicity that $\rho_\f$ has the simple spectrum $0<p_1<p_2<\cdots p_d<1$,
we can restate our result in terms of pdf: a heat $Q_k=\beta^{-1}(\log d+\log p_k)$ is 
released during the process with probability $p_k$.

We obtained a non-generic {\sl atomic} probability distribution: $\Delta Q$ is a
discrete random variable, with each allowed heat quantum 
$Q_k=\beta^{-1}(\log d+\log p_k)$ corresponding to an eigenvalue $p_k$ of 
the final state. According to Eq.~\eqref{EQ-Hsys}, the associated change of the system 
energy is 
$$
E_\f-E_\i=-\beta^{-1}\log p_k-\beta^{-1} \log d + \Delta F.
$$
Energy conservation then implies that the work done on the total system $\cS+\cR$ during 
the transition is equal to the change of the free energy $\Delta F$. A posteriori the 
result is hence interpreted as a fine version of reversibility of the process.

In connection with the erasure process, we further need to consider the limiting case
where $\rho_\f$ becomes a pure state and hence $S_\f\to0$. In this limit,
the probability distribution of $\Delta Q$ acquires extreme outliers  captured by
the singularity of the cumulant generating function
\begin{equation}
\label{1}
\lim_{S_\f\to0}\log\langle\e^{-\alpha\Delta Q}\rangle 
=\begin{cases}
-\frac{\alpha}{\beta}\log d&\mbox{if }\alpha<\beta,\\
0&\mbox{if }\alpha=\beta,\\
\infty&\mbox{if }\alpha>\beta.
\end{cases}
\end{equation}
The first case corresponds to the cumulant generating function of a deterministic
heat dissipation $\Delta Q=\beta^{-1}\Delta S=\beta^{-1}\log d$. In particular, 
we see that $\beta\langle\Delta Q\rangle=\Delta S$ and all higher cumulants vanish. 
The discontinuity at $\alpha=\beta$ and the value of the moment generating function at 
that point is enforced by the fact that $\log\langle\e^{(\beta-\alpha)\Delta Q}\rangle$
is the cumulant generating function of the heat dissipated in the reservoir by the
time reversed evolution. The blow up of the moment generating function for 
$\alpha>\beta$ is the signature of outliers for $\Delta Q<0$.

We believe that the extreme outliers of the heat probability distribution can be
experimentally observed. However,  to see these bumps one
has to look at the whole moment generating function. Moments themselves have no
trace of them. A similar phenomena  of hidden long tails in an adiabatic limit
has been studied in~\cite{CJ}.

We proceed with an extended description of our setup. In particular we define
the dissipated heat through the Full Statistics of the energy change in
the reservoir. We then state the results that allow us to compute the moment
generating function in details.

\subsection{Abstract setup and outline of the heat FS computation}
\label{SEC-Setup}

We consider a finite system $\cS$, with $d$-dimensional Hilbert space, interacting during 
a time interval $[0,T]$ with a reservoir $\cR$ of finite ``size'' $L$. The dynamics of 
the joint system $\cS+\cR$ is governed by the Hamiltonian
\beq
H^{(L)}(t/T)=H_\cS(t/T)+H_\cR^{(L)}+\lambda(t/T)V.
\label{EQ-FullHam}
\eeq
The reservoir Hamiltonian $H^{(L)}_\cR$ and the interaction $V$ are time
independent while the time dependent coupling strength $\lambda(t/T)$ and system 
Hamiltonian $H_\cS(t/T)$ allow us to control the resulting time evolution. In terms of
the rescaled time $s=t/T$, called epoch, the propagator $U^{(T,L)}_s$ associated 
to~\eqref{EQ-FullHam} satisfies the Schr\"{o}dinger equation\footnote{In the whole 
article we choose the time units such that $\hbar=1$.} 
\beq
\frac1T\i\partial_s U^{(T,L)}_s=H^{(L)}(s)U^{(T,L)}_s,\qquad U^{(T,L)}_0=I.
\label{EQ-Evol}
\eeq
In the following we assume that the controls $\lambda(s)$ and $H_\cS(s)$ together 
with their first derivatives are continuous functions of $s\in[0,1]$. More importantly, we 
impose the following boundary conditions: 
\beq
\lambda(0)=\lambda(1)=0,
\label{EQ-lambdabc}
\eeq
which ensure that the system decouples from the environment 
at the initial and the final time, and 
\beq
H_\cS(0)=\beta^{-1}\log d+F_\i,\qquad
H_\cS(1)=-\beta^{-1}\log\rho_\f+F_\f,
\label{EQ-HSbc}
\eeq
where  $F_{\i/\f}$ are arbitrary constants.

The instantaneous thermal equilibrium state at epoch $s$ and inverse temperature $\beta$ 
is
\[
\eta_s^{(L)}=\frac{\e^{-\beta H^{(L)}(s)}}{\tr(\e^{-\beta H^{(L)}(s)})},
\]
which, taking our boundary conditions into account, reduces to
$$
\eta_\i^{(L)}=\eta_0^{(L)}
=\frac Id\otimes\frac{\e^{-\beta H_\cR^{(L)}}}{\tr(\e^{-\beta H_\cR^{(L)}})},\qquad
\eta_\f^{(L)}=\eta_1^{(L)}
=\rho_\f\otimes\frac{\e^{-\beta H_\cR^{(L)}}}{\tr(\e^{-\beta H_\cR^{(L)}})},
$$
at the  initial/final epoch $s=0/1$. The initial state of the joint system is $\eta_\i^{(L)}$, so that its 
state at epoch $s$ is given by 
$$
\rho_s^{(T,L)}=U^{(T,L)}_s\eta_\i^{(L)}U^{(T,L)\ast}_s.
$$
``Local observables'' of the joint system $\cS+\cR$ are operators on $\cH^{(L)}$ which,
for large enough $L$, do not depend on $L$\footnote{We will give a precise definition
of local observables in Section~\ref{sec:AAE thm}.}. We define the thermodynamic limit of 
the instantaneous equilibrium states on local observables by
$$
\eta^{(\infty)}_s(A)=\lim_{L\to\infty}\tr(\eta_s^{(L)}A),
$$
provided the limit on the right hand side exists.

For large $T$, the system's drive is slow: during the long time interval $[0,T]$, the 
variation of the Hamiltonian $H^{(L)}(t/T)$ stays of order $T^0$. The adiabatic evolution 
is obtained by taking the limit $T\to\infty$. The adiabatic theorem for isothermal 
processes~\cite{ASF1,ASF2,JP} states that for any $s\in[0,1]$ and any local observable 
$A$,
\begin{equation}
\label{TLlimit}
\lim_{T \to \infty}\lim_{L \to \infty}\tr(\rho^{(T,L)}_sA)=\eta^{(\infty)}_s(A).
\end{equation}
We will discuss this relation in more details and give precise conditions for its 
validity in Section~\ref{sec:AAE thm}. Here we just note that the order of limits is 
important: one first takes the thermodynamic limit $L\to\infty$ and then the adiabatic 
limit $T\to\infty$.

We identify the dissipated heat $\Delta Q$ with the change of energy in the reservoir as follows. 
Let $P_e^{(L)}$ denote the orthogonal projection on the eigenspace associated to  the 
eigenvalue $e$ of $H^{(L)}_\cR$. The measurement of $H^{(L)}_\cR$ at the initial epoch 
$s=0$ gives $e$ with a probability $\tr(P_e^{(L)}\eta_\i^{(L)})$. After this measurement 
the system  is in the projected state
\[
\frac{P_e^{(L)}\eta_\i^{(L)}P_e^{(L)}}{\tr\left(P_e^{(L)}\eta_\i^{(L)}\right)}.
\]
The second measurement of $H_\cR^{(L)}$ at the final epoch $s=1$, after the system 
undergoes the transformation described by the propagator $U^{(T,L)}_1$, gives 
$e^\prime$ with the probability 
\[
\frac{\tr\left(P_{e^\prime}^{(L)}U^{(T,L)}_1P_e^{(L)}
\eta_\i^{(L)}P_e^{(L)}U^{(T,L)\ast}_1\right)}{\tr\left(P_e^{(L)}\eta_\i^{(L)}\right)}.
\]
It follows that, in this measurement protocol, the probability of observing an amount of
heat $\Delta Q$ dissipated  in the reservoir is
\[
{\mathbb P}^{(T,L)}(\Delta Q)=\sum_{e^\prime-e =\Delta Q}
\tr\left(P_{e^\prime}^{(L)}U^{(T,L)}_1P_e^{(L)}
\eta_\i^{(L)}P_e^{(L)}U^{(T,L)\ast}_1\right).
\]
This distribution is the Full Statistics of the heat dissipation.
Its cumulant generating function is  
$$
\chi^{(T,L)}(\alpha)
=\log\sum_{\Delta Q}\mathbb P^{(T,L)}(\Delta Q)\e^{-\alpha\Delta Q}
=\log\tr\left(\e^{-\alpha H_\cR^{(L)}}U^{(T,L)}_1\e^{\alpha H_\cR^{(L)}}
\eta_\i^{(L)}U^{(T,L)\ast}_1\right).
$$
In view of the boundary conditions~\eqref{EQ-lambdabc} and~\eqref{EQ-HSbc} at the epoch $s=0$,
the reservoir Hamiltonian $H_\cR^{(L)}$ and the full initial Hamiltonian $H^{(L)}(0)$ 
differ by a constant. Thus, we can replace the former by the latter in the 
above relation. Since the   relative R\'enyi $\alpha$-entropy of two states $\rho$, $\sigma$ is defined by
$$
S_\alpha(\rho|\sigma)=\log\tr(\rho^\alpha\sigma^{1-\alpha}),
$$
a simple calculation leads to the identification
\beq
\chi^{(T,L)}(\alpha)=S_{\frac\alpha\beta}(\eta_\i^{(L)}|\rho^{(T,L)}_1).
\label{EQ-RenyiForm}
\eeq
The existence of the thermodynamic limit of Renyi's entropy~\cite{JOPP} implies that
of the FS. Using the adiabatic limit~(\ref{TLlimit}) we obtain
$$
\lim_{T\to\infty}\lim_{L\to\infty}\chi^{(T,L)}(\alpha)
=S_{\frac\alpha\beta}(\eta^{(\infty)}_\i|\eta^{(\infty)}_\f)
= S_{\frac\alpha\beta}(\rho_\i|\rho_\f).
$$
The second equality follows from the boundary condition~\eqref{EQ-lambdabc}: 
the states $\eta_{\i/\f}^{(\infty)}$ factorize and their relative entropy is the
sum of the relative entropies of each factor. Since the initial and the final 
state of the reservoir are identical, their relative entropy vanishes
and we are left with the relative entropy between the initial and final states of the 
system $\cS$ alone. Substituting  $\rho_\i=I/d$ we recover Eq.~\eqref{mainresult}.

A condition regarding the stability of thermal equilibrium states is the main assumption 
required for the validity of Eq.~\eqref{TLlimit} and our analysis in general
(see Section~\ref{sec:AAE thm}). Although this condition is expected to 
hold in a wide class of systems, it is notoriously difficult to prove from basic 
principles. Spin networks, and generic spin-fermion models are among the relevant 
systems for which the condition has been rigorously 
established 
although most often with a large temperature--weak
coupling assumption. We specialize our discussion to one of these models, in
which the thermal states are known to be stable for a weak enough interaction.
It describes a one dimensional fermionic chain with an impurity. We would like
to stress that this choice of model is not central to the results presented
here. They hold for any model exhibiting the same stability behaviour at
equilibrium.

\subsection{A fermionic impurity model}
\label{SEC-model}

\begin{figure}
\begin{center}
\includegraphics[width=.8\textwidth]{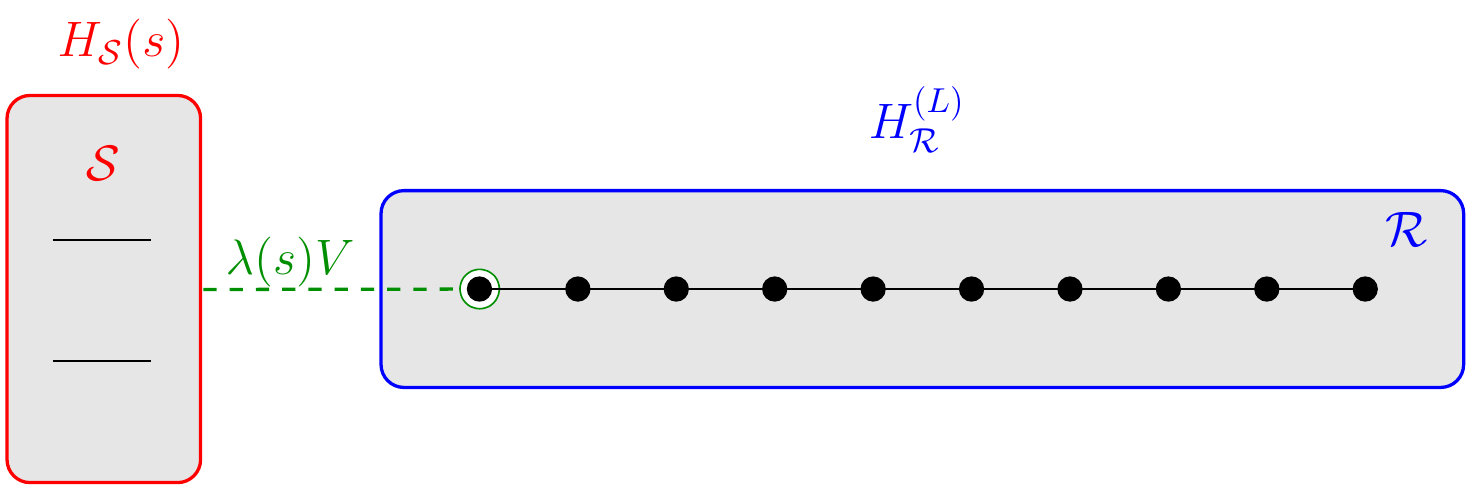}
\end{center}
\caption{A $2$--level quantum system $\cS$ interacts with a gas of fermions $\cR$ on a one
dimensional lattice of size $L$. The epoch-dependent interaction $\lambda(s)V$ couples
$\cS$ to the fermions on the first site of this lattice.}
\label{FIG-spinF}
\end{figure}

We describe a concrete realization of the abstract setup of the previous
section. A 2-level quantum system $\cS$ interacts with a reservoir
$\cR$, a gas of spinless  fermions on a one dimensional lattice of size
$L$ (see Figure~\ref{FIG-spinF}). The system and the reservoir are coupled by a
dipolar rotating-wave type interaction between $\cS$ and the fermions on the
first site of the lattice. When uncoupled, the reservoir is a free Fermi gas in
thermal equilibrium at inverse temperature $\beta$. The thermodynamic limit is
obtained by taking the size $L$ of the lattice to infinity. The details are as follows.

The lattice sites are labeled by $x\in\Lambda^{(L)}=\{1,2,\ldots,L\}$. The one-particle 
Hilbert space of the reservoir is $\ell^2(\Lambda^{(L)})$ and we denote by
$\delta_x$ the delta-function at site $x$. The reservoir is thus described by the 
antisymmetric Fock space $\cH_\cR^{(L)}=\Gamma(\ell^2(\Lambda^{(L)}))$,
a $2^L$-dimensional Hilbert space. The creation/annihilation operator for a fermion at site $x\in\Lambda^{(L)}$ is $c^\ast (x)$/$c(x)$.
These operators obey the canonical anti-commutation relation
$$
\{c(x),c^\ast(x')\}
=c(x)c^\ast(x')+c^\ast (x')c(x)
=\delta_{x,x'}.
$$
The reservoir Hamiltonian
\[
H_\cR^{(L)}=\kappa \sum_{\substack{x,y\in\Lambda^{(L)}\\|x-y|=1 }} 
c^\ast(x)c(y)
\]
is the second quantization of $\kappa \Delta^{(L)}$, where $\Delta^{(L)}$ is  the discrete Laplacian on $\Lambda^{(L)}$ with Dirichlet 
boundary conditions,
$$
(\Delta^{(L)}f)(x)
=\begin{cases}
f(2)&\mbox{for }x=1,\\
f(x+1)+f(x-1)&\mbox{for }1<x<L,\\
f(L-1)&\mbox{for }x=L.\\
\end{cases}
$$
Thus, $H_\cR^{(L)}$ corresponds to homogeneous hopping between neighboring lattice sites 
with a hopping constant $\kappa>0$.

The Hilbert space of the system $\cS$ is $\cH_\cS=\cc^2$. We denote by $\sigma_x$,
$\sigma_y$ and $\sigma_z$ the usual Pauli matrices on $\cH_\cS$.
In view of the initial condition $\rho_\i=I/2$ and boundary conditions~\eqref{EQ-HSbc}, 
we can assume, without loss of generality, that its Hamiltonian is given by
\beq 
H_\cS(s)=\epsilon(s)I+\gamma(s)\sigma_z.
\label{ham-s}
\eeq
The total Hilbert space is $\cH^{(L)}=\cH_\cS\otimes\cH^{(L)}_\cR$.
The coupling is achieved by a rotating-wave type interaction between ${\cal S}$ and 
the fermion on the first lattice site 
\[
V=\sigma^- \otimes c^*(1) + \sigma^+ \otimes c(1),
\]
where $\sigma^{\pm}=\frac{1}{2}(\sigma_x\pm \i \sigma_y)$. Note that $V$ is a local 
observable: it does not depend on the lattice size $L$. This restriction is not strictly necessary but we will not elaborate on this point here. 

The Jordan-Wigner transformation maps the fermionic impurity 
model to a free Fermi gas with one-particle Hilbert space 
$\cc\oplus\ell^2(\Lambda^{(L)})$ and one-particle Hamiltonian of the 
Friedrich's type 
\beq
h(s)=(\epsilon(s)+\gamma(s))I-2\gamma(s)|1\rangle \langle 1|
-\lambda(s)(|1\rangle\langle \delta_1| + |\delta_1\rangle \langle 1|) + 
\kappa \Delta^{(L)},
\label{fri}
\eeq
where $|1\rangle$ denotes the basis vector of $\cc$.
This allows for  a detailed study of the   mathematical and physical aspects of this model;  see  \cite{AJPP1, JKP}.

\section{Adiabatic limits for thermal states}
\label{SEC-AdiabaticLimit}

This section starts with a discussion of the relevant time-scales of the fermionic 
impurity model of Section~\ref{SEC-model}. Then, we investigate the various adiabatic
regimes that can be reached by appropriate separations of these time-scales. In 
particular, we explain why the order of limits in Eq.~\eqref{TLlimit} is relevant
for the realization of a quasi-static erasure protocol.

\subsection{Time-scales in the impurity model}
\label{sec:time scales}

Adiabatic theory provides a tool to study the dynamics of systems which feature
separation of some relevant physical time-scales. To elucidate its meaning in our
setup we compare the adiabatic time $T$ with the three relevant dynamical time-scales
of our model. We discuss the three adiabatic theorems corresponding to 
different ordering of $T$ with respect to these time-scales.

For each fixed epoch $s$ we consider the time-scales associated to the dynamics
generated by the instantaneous Hamiltonian 
$H^{(L)}(s)=H_\cS(s)+\lambda(s)V+ H_\cR^{(L)}$. In the following discussion we assume 
that for $s\in]0,1[$ the $s$-dependence of these time-scales is negligible and we omit 
the variable $s$ from our notation. We reinstate the $s$-dependence in the last paragraph 
of this subsection.

\begin{description}
\item[{\bf $T_\cS$: the recurrence time of $\cS$.}]This is the quantum analogue of the 
Poincaré recurrence time, the time after which the isolated ($\lambda=0$) system $\cS$ 
returns to nearly  its initial state; see \cite{BM, CV}. For typical initial states, this time is 
inversely proportional to the mean level spacing  of the system Hamiltonian 
$H_\cS$. For the fermionic impurity model described in the previous section one has 
$$
T_\cS\sim \frac{1}{\gamma}.
$$
We recall that we use physical units in which energy is the inverse of time, and 
hence $T_\cS$ is indeed a time-scale.

\item[{\bf $T_{\cS+\cR}$: the recurrence time of $\cS+\cR$.}]The same as $T_\cS$, but
for the coupled  ($\lambda\neq 0$) system $\cS+\cR$. The eigenvalues of the discrete
Laplacian $\Delta^{(L)}$ are
$$
\varepsilon_k=2\cos\left(\frac{k\pi}{L+1}\right),\quad (k=1,\ldots,L),
$$
and those $H_\cR^{(L)}$ are
$$
\kappa \sum_{k=1}^L n_k\varepsilon_k,\quad(n_k=0,1,2).
$$
It follows that the diameter of the spectrum of $H_\cR^{(L)}$ is $\cO(L)$ for large $L$. The same is true
for the full Hamiltonian $H^{(L)}$, while $\dim(\cH^{(L)})=2^{L+1}$. 
Thus, the mean level spacing of $H^{(L)}$ is $\cO(L 2^{-(L+1)})$ and we conclude that
$$
T_{\cS+\cR}=\frac1{\cO(L\,2^{-(L+1)})}
$$
diverges in the the thermodynamic limit $L\to\infty$.

\item[{\bf $T_{m}$: the equilibration time.}]  This is the time needed for the coupled
system $\cS+\cR$ to return to thermal (quasi--)equilibrium after a localized
perturbation. In the thermodynamic limit $L=\infty$, the system remains in
thermal equilibrium after this time which, in this case, coincides with the 
mixing time. However, for finite $L$, recurrences appear for 
times of order $T_{\cS+\cR}$ which is much larger than $T_m$. In
Section~\ref{sec:AAE thm} we shall argue that for small enough $\lambda>0$,
$T_{m}$ stays finite as $L\to\infty$.

In the weak coupling regime, Fermi's golden rule gives the dependence
$T_{m}(\lambda)=\cO(\lambda^{-2})$ on the interaction strength $\lambda$. Note
in particular that $T_{m}(0)=\infty$. Equilibration is not possible without 
interaction between $\cS$ and $\cR$.
\end{description}

\begin{figure}[h]
\begin{center}
\includegraphics[width=.55\textwidth]{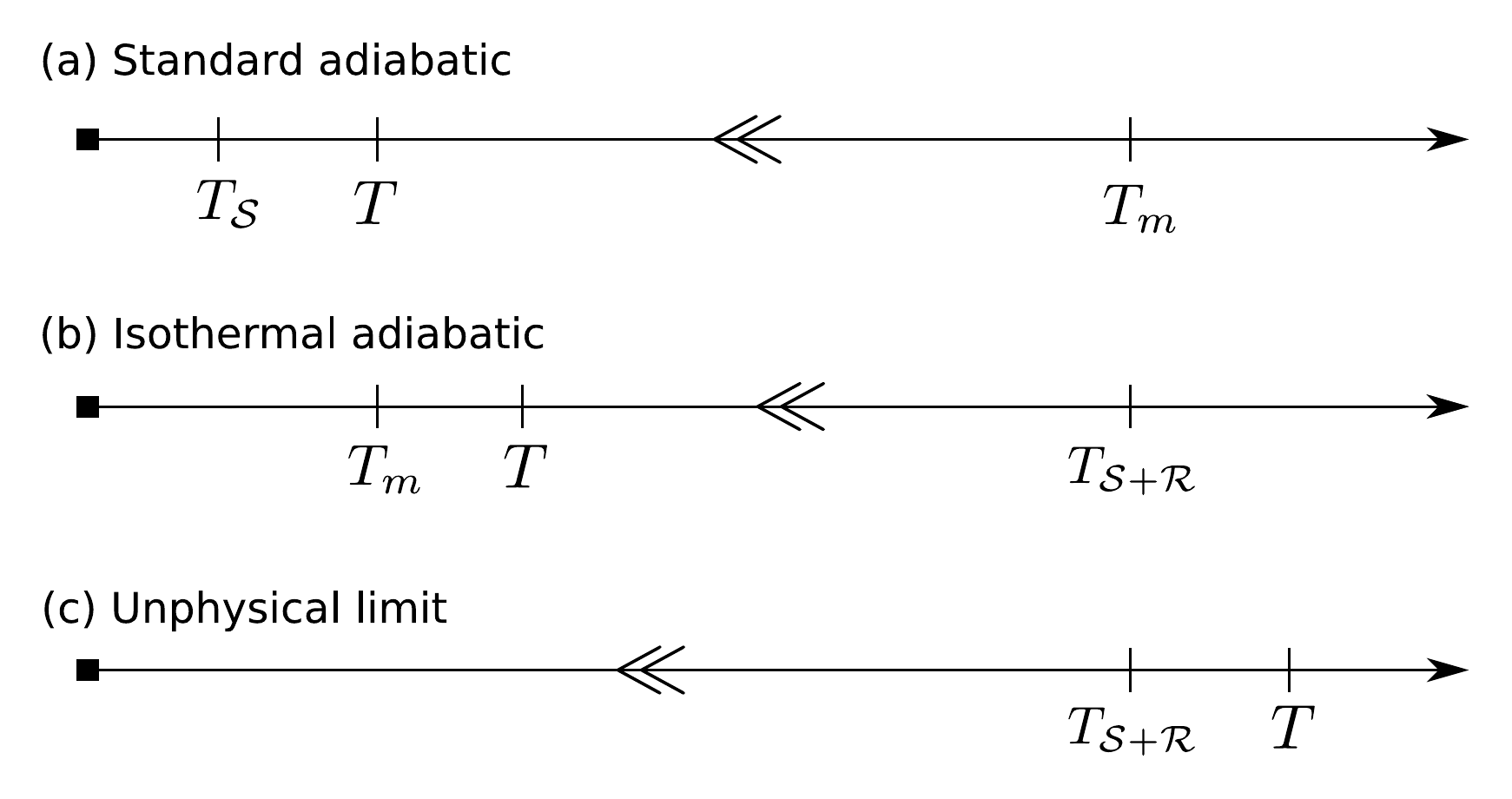}
\end{center}
\caption{\label{fig:time scales} The different time-scale orderings of the
adiabatic theorems. In the ordering (a) of the standard adiabatic theorem,
the relevant evolution is that of the system $\cS$ with its associated
time-scale $T_\cS$. The ordering (c) is associated to the isothermal adiabatic
theorem. The ordering for the weak coupling adiabatic theorem (b) is similar with
the constraint $T\sim\lambda^{-2}$. In both cases  the
relevant time-scale is the thermalization time $T_m$. The ordering (d) corresponds to the
unphysical limit where $T$ is taken to infinity before the thermodynamic limit.}
\end{figure}

In the physical systems we have in mind, these time-scales are naturally ordered
as 
$$
T_\cS\ll T_{m}\ll T_{\cS+\cR}.
$$
Three physically relevant regimes and one unphysical adiabatic regime are consistent with this 
ordering (see Figure~\ref{fig:time scales}):
\begin{enumerate}
\item $T_\cS\ll T\ll T_{m}\ll T_{\cS+\cR}$ (Figure~\ref{fig:time scales}~(a)). The 
adiabatic regime is reached by taking first $L\to\infty$, then $\lambda\to0$, and finally 
$T\to\infty$. After the $\lambda\to0$ limit, the system $\cS$ decouples from the 
reservoir and the $T\to\infty$ limit yields the standard adiabatic theorem of quantum 
mechanics~\cite{BF,Ka} applied to the isolated system $\cS$.

\item $T_\cS\ll T\sim T_{m}\ll T_{\cS+\cR}$ (Figure~\ref{fig:time scales}~(b)). The
ordering is the same as in the previous case, but $T$ and $T_m$ remain comparable. To 
reach this regime, one first take $L\to\infty$, and then simultaneously $\lambda\to0$
and $T\sim\lambda^{-2}\to\infty$. This procedure yields the weak coupling 
adiabatic theorem of Davies and Spohn~\cite{DS,TAS}.

\item $T_\cS<T_m\ll T\ll T_{\cS+\cR}$ (Figure~\ref{fig:time scales}~(c)). This regime
corresponds to first taking $L\to\infty$, and then $T\to\infty$ keeping 
$\lambda\sim\cO(1)$. It is controlled by the adiabatic theorem for isothermal 
processes~\cite{ASF1,ASF2,JP}. After tracing out the degrees of freedom of the reservoir this 
regime should be equivalent to the Markovian adiabatic theory~\cite{SL,Jo,AFGG}.

\item $T_{\cS+\cR}\ll T$ (Figure~\ref{fig:time scales}~(d)). This unphysical regime is
reached by first taking $T\to\infty$. The standard adiabatic theorem applies again, but
this time to the joint system $\cS+\cR$. The subsequent thermodynamic limit $L\to\infty$ 
enforces an infinitely slow driving. We devote the following
section to show that the superhero\footnote{We heard a rumor that in the
upcoming X-men movie there would be a new character with a superpower that
allows her to wait infinitely long.} adiabatic theorem associated to this regime
gives very different predictions compared to the isothermal adiabatic theorem.
\end{enumerate}

\noindent{\bf Remark.}
The family of Hamiltonians $\{H^{(L)}(s)\}_{s\in[0,1]}$ might possess
exceptional points at which one or more of the above time-scales diverge. In the
standard adiabatic theory these exceptional points correspond to eigenvalue
crossings, i.e., accidental degeneracies. The zeroth order adiabatic
approximation still holds in the presence of finitely many such exceptional
points. In the isothermal adiabatic theory, exceptional points occur whenever
$\lambda(s)=0$. Similar to the standard theory, the adiabatic approximation
holds also in the presence of finitely many such points. Note in particular that the 
erasure process has exceptional  points at the initial/final epoch $s=0/1$.

\subsection{The adiabatic limit for thermal states at finite $L$}

Let us apply the standard adiabatic theorem~\cite{BF,Ka} to the full Hamiltonian 
$H^{(L)}(s)$ for finite $L$. For simplicity, we assume that the family 
$\{H^{(L)}(s)\}_{s\in]0,1[}$ has no exceptional points and admits the representation
$$
H^{(L)}(s)=\sum_{k}e_k(s)P_k(s),
$$
where the projections $P_k(s)$ are continuously differentiable functions of $s$. 
Then the adiabatic theorem states that
\[
\lim_{T\to\infty}U^{(T,L)}_sP_k(0)U^{(T,L)\ast}_s=P_k(s).
\]
Hence, given the initial state
\[
\eta_\i^{(L)}=\frac{\e^{-\beta H^{(L)}(0)}}{\tr(\e^{-\beta H^{(L)}(0)})}
=\frac1{Z_0^{(L)}}\sum_{k}\e^{-\beta e_k(0)}P_k(0),
\]
the final state $\rho^{(T,L)}_1$ satisfies
$$
\lim_{T\to\infty}\rho^{(T,L)}_1
=\lim_{T\to\infty}U^{(T,L)}_1\eta_\i^{(L)}U^{(T,L)\ast}_1
=\frac1{Z_0^{(L)}}\sum_{k}\e^{-\beta e_k(0)}P_k(1),
$$
which only coincides with $\eta_\f^{(L)}$ if
$Z_0^{(L)}\e^{-\beta e_k(1)}=Z_1^{(L)}\e^{-\beta e_k(0)}$ for all $k$. This is of course 
a very strong constraint which, in particular, is not satisfied in an erasure protocol.

\subsection{The isothermal adiabatic theorem}
\label{sec:AAE thm}

The main purpose of this section is to formulate a precise statement of the
isothermal adiabatic theorem, which is the main technical ingredient of our analysis
of quasi-static erasure processes. This requires some preparation and 
we will start by discussing the thermodynamic limit
$L\to\infty$, and in particular the fate of families $\{\rho^{(L)}\}_{L\ge0}$ of finite
volume states in this limit. Then we will introduce  the notion of {\em ergodicity} which is the main dynamical assumption of
the isothermal adiabatic theorem.

\paragraph{The thermodynamic limit.} To avoid
technically involved algebraic techniques, we will only work with a {\sl potential}
infinity, i.e., all infinite volume objects will be defined as limits of
their finite volume counterparts. A drawback of this approach is that we cannot
explain the proof of the isothermal adiabatic theorem, Eq.\eqref{TLlimit}, in details. 
This proof, which requires the algebraic machinery of quantum statistical mechanics is, 
however, available in the existing literature~\cite{ASF1,ASF2,JP} and it is also given in the
companion paper~\cite{BFJP}. In the following, we denote all infinite volume quantities 
with the superscript ${}^{(\infty)}$.

A central role in the definition of the thermodynamic limit is played by the
set $\cA$ of so-called \emph{local observables} of the infinite volume system
$\cS+\cR$. For our purposes, it will suffice to consider $\cA=\cup_{L\ge0}\cA^{(L)}$,
where $\cA^{(L)}$ is the set of operators which are finite sums of monomials of the form
$$
D\otimes c^\ast(x_1)\cdots c^\ast(x_n)
c(y_1)\cdots c(y_m),\qquad 
(x_i,y_i\in\Lambda^{(L)}),
$$
where $D$ is an operator on $\cH_\cS$. Note that $\cA^{(L)}\subset\cA^{(L')}$
whenever $L\le L'$. By definition, operators in $\cA$ involve only a
finite number of lattice sites of the Fermi gas $\cR$ and hence remain well defined as
operators on $\cH^{(L)}$ for large enough but finite $L$. In fact, $\cA^{(L)}$ coincides
with the set of all operators on $\cH^{(L)}$. In particular, sums and products of 
elements of $\cA$ are themselves elements of $\cA$ (i.e., $\cA$ is an algebra).

Assume that for each $L\ge0$, $\rho^{(L)}$ is a density matrix on $\cH^{(L)}$. Given
a local observable $A\in\cA$, the expectation $\rho^{(L)}(A)=\tr(\rho^{(L)}A)$ is
well defined for large enough $L$. We say that the sequence $\{\rho^{(L)}\}_{L\ge0}$
has the thermodynamic limit $\rho^{(\infty)}$ whenever, for each $A\in\cA$, the limit
$$
\rho^{(\infty)}(A)=\lim_{L\to\infty}\tr(\rho^{(L)}A)
$$
exists. We remark that there may be no density matrix on $\cH^{(\infty)}$
such that $\rho^{(\infty)}(A)=\tr(\rho^{(\infty)}A)$. Nevertheless, the infinite
volume state $\rho^{(\infty)}$ defined in this way provides an expectation functional
on $\cA$ with the properties $\rho^{(\infty)}(I)=1$ and
$0\le\rho^{(\infty)}(A^\ast A)\le\|A\|^2$ for all $A\in\cA$.

We also note that $H_\cR^{(\infty)}$, the energy of the infinite reservoir, {\sl is not} 
a local observable and therefore need not have a finite expectation in a thermodynamic
limit state $\rho^{(\infty)}$. This is physically consistent with the fact that
$\rho^{(\infty)}$ may describe a state of the infinite system with infinite energy
(this will indeed be the case for all the thermodynamic limit states relevant to our
analysis of erasure processes).
On the contrary, $H_\cS$ {\sl is} a local observable and the energy of the
system $\cS$ has finite expectation in any thermodynamic limit state.

Assume now that for each $L\ge0$, besides the state $\rho^{(L)}$, we also have a
unitary propagator $U^{(L)}_t$ for the finite system $\cS+\cR$. 
Since $U^{(L)}_t\in\cA^{(L)}$, for any $A\in\cA$ we have 
$U^{(L)\ast}_t AU^{(L)}_t\in\cA^{(L)}$ for large enough $L$ so that 
$\tr(\rho^{(L)}U^{(L)\ast}_t AU^{(L)}_t)$ is well defined. We shall
say that the sequence $\{U^{(L)}_t\}_{L\ge0}$ defines a dynamics for $\rho^{(\infty)}$ on 
the time interval ${\cal I}$ if
$$
\rho^{(\infty)}_t(A)
=\lim_{L\to\infty}\tr\left(\rho^{(L)}U^{(L)\ast}_t AU^{(L)}_t\right)
$$
exists for all $A\in\cA$ and all $t\in {\cal I}$. Note that the existence of this
limiting dynamics depends not only on the sequence of finite volume propagators,
but also on the sequence of finite volume states.

Decades of effort were devoted by the theoretical and mathematical physics communities
to the construction and characterization of thermodynamic limit states of quantum systems
and their dynamics. We refer the reader to~\cite{Ru,BR1,BR2} for detailed expositions of 
the resulting theory.

Specializing to our impurity model, for each epoch $s$, the
instantaneous thermal state $\eta^{(L)}_s$ admits a thermodynamic limit
$\eta^{(\infty)}_s$. Equally importantly for our problem, the propagators
$U^{(T,L)}_s$ define a dynamics for these states and in particular
$$
\rho^{(T,\infty)}_s(A)=\lim_{L\to\infty}\tr\left(\rho^{(T,L)}_s A\right)
=\lim_{L\to\infty}\tr\left(\eta_\i^{(L)}U^{(T,L)\ast}_sAU^{(T,L)}_s\right)
$$
exists for all $A\in\cA$ and $s\in[0,1]$.

\paragraph{Ergodicity.}
As already mentioned in Section~\ref{sec:time scales}, the adiabatic theory of
isothermal processes requires the instantaneous dynamics at each fixed epoch $s$ 
(with the possible exception of finitely many of them) to have the property that
a local perturbations of the instantaneous thermal equilibrium state should relax 
to this equilibrium state. We now give a more precise statement of this requirement
in terms of  the  ergodic property of the instantaneous dynamics. 

Let $\{\rho^{(L)}\}_{L\ge0}$ be a sequence of finite volume states with
thermodynamic limit $\rho^{(\infty)}$. For any non-zero  $B\in\cA$, the perturbed states
$$
\rho^{(L)}_B=\frac{B^\ast \rho^{(L)}B}{\tr(B^\ast \rho^{(L)}B)}
$$
are well defined for large enough $L$. Using the cyclic property of the trace, one easily 
shows that
$$
\left|\tr\left(\rho^{(L)}BAB^\ast\right)\right|\le\|A\|\tr\left(B^\ast\rho^{(L)}B\right).
$$ 
Thus, the thermodynamic limit
$$
\rho_B^{(\infty)}(A)
=\lim_{L\to\infty}\tr\left(\rho_B^{(L)}A\right)
=\lim_{L\to\infty}\frac{\tr\left(\rho^{(L)}BAB^\ast\right)}{\tr(\rho^{(L)}BB^\ast)}
=\frac{\rho^{(\infty)}(BAB^\ast)}{\rho^{(\infty)}(BB^\ast)}
$$
also exists and defines a {\sl local perturbation} $\rho_B^{(\infty)}$ of the 
state $\rho^{(\infty)}$. Assume that the sequence of Hamiltonians
$\{H^{(L)}\}_{L\ge0}$ defines a dynamics
$$
\rho_{B,t}^{(\infty)}(A)=\lim_{L\to\infty}\left(\rho^{(L)}_B
\e^{\i tH^{(L)}}A\e^{-\i tH^{(L)}}\right)
$$
on these states. The state $\rho^{(\infty)}$ is said to be ergodic with respect to this 
dynamics if, for all $A,B\in\cA$, we have 
$$
\lim_{t\to\infty}\frac1{2t}\int_{-t}^t 
\rho^{(\infty)}_{B,u}\left(A\right)\d u=\rho^{(\infty)}(A).
$$
Note that it follows from this relation that $\rho^{(\infty)}$ is invariant under the
dynamics, i.e., that 
$$
\rho^{(\infty)}_t(A)=\lim_{L\to\infty}\tr\left(\rho^{(L)}
\e^{\i tH^{(L)}}A\e^{-\i tH^{(L)}}\right)=\rho^{(\infty)}(A)
$$
for all $t\in\rr$ and $A\in\cA$.

Ergodicity, i.e., return to equilibrium for autonomous dynamics, has been proven for a 
large number of physically relevant 
models~\cite{BoM,AM,JP1,BFS,JP2,DJ,FMSU,FMU,FM,AJPP1,JOP1,AJPP2,JOP2,MMS1,MMS2,dRK}. 
In the case of our impurity model, ergodicity of the instantaneous thermal state 
$\rho^{(\infty)}_s$  with respect to the instantaneous dynamics generated by the 
Hamiltonians $H_s^{(L)}$ holds for small enough  $\lambda(s)\neq 0$ assuming that the coupling between 
${\cal S}$ and ${\cal R}$ is effective, i.e., 
\[
2\gamma(s)\in]-2\kappa,2\kappa[, 
\]
where $[-2\kappa,2\kappa]=\sp(\kappa \Delta)$, $\Delta=\lim_{L\rightarrow \infty}\Delta^{(L)}$
being the half-line discrete Laplacian; see~\cite{AJPP1, JKP}.

We are now ready to state the adiabatic theorem that leads to our
results. By the discussion above the assumptions of the theorem can be satisfied
in our impurity model by an appropriate choice of $\kappa$ and the coupling strength $\lambda(s)$. The same applies 
to the choice of the boundary conditions  (\ref{EQ-HSbc}), since one  may assume from the outset that the final state $\rho_{\rm f}$ is described 
by a diagonal density matrices on $\cH_\cS=\cc^2$. 
\bet\label{THM-IsoThermal}
Assume that at any epochs $0<s<1$, the thermal state $\eta_s^{(\infty)}$ is ergodic 
with respect to the dynamics generated by the sequence of Hamiltonian 
$\{H^{(L)}(s)\}_{L\ge0}$. 
Assume also that $H_\cS(s)$ and $\lambda(s)$  are
continuously differentiable in $s$ on $[0,1]$. Then, in the limit $T\to\infty$,
the state $\eta_\i^{(\infty)}$ evolves along the path of instantaneous thermal equilibrium
states at the fixed inverse temperature $\beta$,
\beq\label{eq-Adiabatic Limit}
\lim_{T\to\infty}\sup_{A\in\cA,\|A\|=1}
\big|\rho^{(T,\infty)}_s(A)-\eta_s^{(\infty)}(A)\big|=0,
	\eeq
for every $s\in [0,1]$. In the adiabatic limit, the evolution is hence a quasi--static 
isothermal process.
\eet
The theorem has been proved in  \cite{ASF1, ASF2, JP}. The 
proof uses Araki's perturbation theory and the adiabatic theorem without gap
condition~\cite{AE,Te}. The crucial result of the former is that all the
instantaneous thermal equilibrium states $\eta^{(\infty)}_s$ are mutually
quasi--equivalent, and can be represented as vectors in the same GNS
representation (i.e., in the same Hilbert space). In this representation,
the dynamics is governed by a time-dependent standard Liouvilian 
$\mathcal{L}_s$. If the instantaneous dynamics at a given epoch $s$ is ergodic,
then $0$ is a simple eigenvalue of $\mathcal{L}_s$ and the vector
representative of $\eta^{(\infty)}_s$ is the corresponding eigenvector. Since
$\mathcal{L}_s$ inherits the differentiability properties of the finite volume
Hamiltonians  $H^{(L)}(s)$, the adiabatic theorem without gap condition implies
the above theorem. We now move on to discuss its consequences.

\medskip
\noindent{\bf Remark.} Our analysis of erasure processes can easily be generalized to
a wider class of models. However, these generalizations are restricted to
thermal states of the joint system $\cS+\cR$ describing pure thermodynamic phases. 
We particularly  emphasize that our results do not apply to adiabatic phase transition 
crossing.

\section{Heat full statistics in the adiabatic limit}
\label{SEC-FCS}

The purpose of this section is to derive the Full Statistics of the
heat dissipated into the reservoir during the quasi-static process described in
the introduction. We start the section with a detailed discussion of the
energy balance and its thermodynamic limit. Then, starting with 
Relation~\eqref{EQ-RenyiForm}, we study the thermodynamic limit of the heat FS
and, invoking Theorem~\ref{THM-IsoThermal}, its adiabatic limit.

For finite $L$ and $T$, the expected value of the work done on the joint system
$\cS+\cR$ during the state transition $\rho^{(T,L)}_0\to\rho^{(T,L)}_1$ mediated by
the propagator $U_1^{(T,L)}$ is given by 
\beq
W^{(T,L)}=\tr\left(\rho_1^{(T,L)}H^{(L)}(1)\right)-\tr\left(\rho_0^{(T,L)}H^{(L)}(0)\right)
\label{EQ-WForm}.
\eeq
We have 
\[
\begin{split}
W^{(T,L)}&=\int_0^1\partial_s\tr\left(\rho_s^{(T,L)}H^{(L)}(s)\right)\d s\\
&=\int_0^1\partial_s\tr\left(\eta_\i^{(L)}U_s^{(T,L)\ast}
H^{(L)}(s)U_s^{(T,L)}\right)\d s\\
&=\int_0^1\tr\left(\eta_\i^{(L)}U_s^{(T,L)\ast}\left(\i T[H^{(L)}(s),H^{(L)}(s)]+
\partial_s H^{(L)}(s)\right)U_s^{(T,L)}
\right)\d s\\
&=\int_0^1\tr\left(\rho_s^{(T,L)}(\dot{H}_\cS(s)+\dot{\lambda}(s)V)\right)\d s,
\end{split}
\]
where we have used the evolution equation~\eqref{EQ-Evol}. 

The expected value of the
change in the energy of the system $\cS$ is
\beq
\langle\Delta E_\cS^{(T,L)}\rangle
=\tr\left(\rho_1^{(T,L)}H_\cS(1)\right)-\tr\left(\rho_0^{(T,L)}H_\cS(0)\right).
\label{EQ-DeltaEForm}
\eeq
Finally, the expected value of the change in the reservoir energy is 
$$
\langle\Delta Q^{(T,L)}\rangle
=\tr\left(\rho_1^{(T,L)}H_\cR^{(L)}\right)
-\tr\left(\rho_0^{(T,L)}H_\cR^{(L)}\right).
$$ 
Although the individual terms on the right hand side of the last identity do not admit a 
thermodynamic limit, their difference remain well defined in the limit $L\to\infty$. 
This becomes clear when writing the first law
$$
\langle\Delta Q^{(T,L)}\rangle=W^{(T,L)}-\langle\Delta E_\cS^{(T,L)}\rangle,
$$
which obviously follows from~\eqref{EQ-WForm}, \eqref{EQ-DeltaEForm} and the
boundary condition~\eqref{EQ-lambdabc}. Indeed, both
$$
W^{(T,\infty)}
=\lim_{L\to\infty}W^{(T,L)}
=\int_0^1\rho_s^{(T,\infty)}\left(\dot{H}_\cS(s)+\dot{\lambda}(s)V\right)\d s,
$$
and
$$
\langle\Delta E_\cS^{(T,\infty)}\rangle
=\lim_{L\to\infty}\langle\Delta E_\cS^{(T,L)}\rangle
=\rho_1^{(T,\infty)}\left(H_\cS(1)\right)-\rho_0^{(T,\infty)}\left(H_\cS(0)\right),
$$
are well defined. 

In the adiabatic limit $T\to\infty$, the work done on the joint system coincides with 
the increase of its free energy: Duhamel's formula and Theorem~\ref{THM-IsoThermal} 
yield
\begin{align*}
\lim_{T\to\infty}W^{(T,\infty)}
&=\int_0^1\eta^{(\infty)}_s(\dot{H}_\cS(s)+\dot{\lambda}(s)V)\d s\\
&=\lim_{L\to\infty}\int_0^1\frac{\tr(\e^{-\beta H^{(L)}_s}
(\dot{H}_\cS(s)+\dot{\lambda}(s)V))}
{\tr(\e^{-\beta H^{(L)}_s})}\d s\\
&=-\lim_{L\to\infty}\frac1\beta\int_0^1\partial_s\log\tr(\e^{-\beta H^{(L)}_s})\d s\\
&=-\frac{1}{\beta}\log\tr(\e^{-\beta H_\cS(1)})
+\frac1\beta\log\tr(\e^{-\beta H_\cS(0)})\\
&=F_\f-F_\i=\Delta F.
\end{align*}
The equality between  work and free energy is the signature of a reversible
process: the work done can be recovered from the system by reversing the
trajectory. Recalling from classical thermodynamics that for isothermal
processes we have 
\[\Delta F-\langle\Delta Q\rangle=W-\beta^{-1}\Delta S,\]
the equality between work and free energy leads to saturation in the
Landauer bound:
$$
\beta\lim_{T \to \infty}\lim_{L \to \infty}\langle\Delta Q^{(T,L)}\rangle=\Delta S.
$$
As already mentioned in the introduction, a mathematical proof of this saturation
can be obtained using an appropriate microscopic version of the entropy balance 
equation~\cite{JP}.

Using standard algebraic techniques of quantum statistical mechanics, it is fairly easy 
to show that the thermodynamic limit of Renyi's relative entropy for the fermionic 
impurity model
$$
S_\alpha(\eta^{(\infty)}_\i|\rho^{(T,\infty)}_1)
=\lim_{L\to\infty}S_\alpha(\eta^{(L)}_\i|\rho^{(T,L)}_1)
$$
exists. The left hand side of this identity can be expressed in terms of relative modular
operators in the GNS Hilbert space associated to the state $\eta^{(L)}_\i$ 
(see~\cite{JOPP}, a detailed proof can be found in~\cite{BFJP}). This representation 
shows in particular that, as a function of $\alpha$, the entropy 
$S_\alpha(\eta^{(\infty)}_\i|\rho^{(T,\infty)}_1)$
is analytic in the strip $0<\Re{\alpha}<1$ and continuous on its closure.

Recalling Relation~\eqref{EQ-RenyiForm} between R\'enyi's entropy and
cumulant generating function, we can write
\beq\label{eq-Renyi TDL}
\chi^{(T,\infty)}(\alpha)=\lim_{L\to\infty}\chi^{(T,L)}(\alpha)
=S_{\frac\alpha\beta}(\eta^{(\infty)}_\i|\rho_1^{(T,\infty)}),
\eeq
and conclude that the characteristic function (i.e., the Fourier transform) of the heat FS
$$
\varphi^{(T,L)}(\alpha)=\sum_{\Delta Q}\e^{\i\alpha\Delta Q}\,\mathbb{P}^{(T,L)}(\Delta Q)
$$
converges pointwise, for all $\alpha\in\rr$, towards the continuous function
$$
\varphi^{(T,\infty)}(\alpha)
=\e^{S_{-\i\frac\alpha\beta}(\eta^{(\infty)}_\i|\rho_1^{(T,\infty)})}
$$
as $L\to\infty$. Levy's continuity theorem~\cite[Section~1.7]{Bi} allows us to conclude
that for $T>0$, there exists a pdf $\mathbb P^{(T,\infty)}$ which is the 
weak limit of the finite volume pdf $\mathbb P^{(T,L)}$, i.e.,
\[
\lim_{L\to\infty}\sum_{\Delta Q}f(\Delta Q)\mathbb P^{(T,L)}(\Delta Q)
=\int_\rr f(\Delta Q)\d\mathbb P^{(T,\infty)}(\Delta Q)
\]
for any bounded continuous function $f$.

It remains to take the adiabatic limit $T\to\infty$. The uniform convergence 
in~\eqref{eq-Adiabatic Limit} and the properties of relative modular operators acting
on the GNS Hilbert space imply that
$$
\lim_{T\to\infty}S_\alpha(\eta_\i^{(\infty)}|\rho_1^{(T,\infty)})
=S_\alpha(\eta_\i^{(\infty)}|\eta_\f^{(\infty)})
=-\alpha\log d+\log\tr(\rho_\f^{1-\alpha}),
$$
the convergence being uniform on any compact subset of the strip $0\le\Re(\alpha)<1$.
The detailed proof can be found in~\cite{BFJP} (see also~\cite{JPPP} where a similar
argument has been used). Thus, we have obtained the following expression for the 
cumulant generating function of the dissipated heat in the adiabatic limit,
\begin{equation}
\label{FCSg}
\bar{\chi}(\alpha)
=\lim_{T\to\infty}\lim_{L\to\infty}\chi^{(T,L)}(\alpha)
=S_{\frac\alpha\beta}(\eta_\i^{(\infty)}|\eta_\f^{(\infty)})
=-\frac\alpha\beta\log d+\log\tr(\rho_\f^{1-\frac\alpha\beta}),
\end{equation}
which is the result~\eqref{mainresult} stated in the introduction. Since the 
limiting characteristic function
\beq
\bar\varphi(\alpha)=\lim_{T\to\infty}\varphi^{(T,\infty)}(\alpha)
=\e^{\bar{\chi}(-\i\alpha)}
=d^{\i\frac\alpha\beta}\tr\left(\rho_\f^{1+\i\frac\alpha\beta}\right)
\label{EQ-LimCharF}
\eeq
is continuous at $\alpha=0$, we can again invoke Levy's continuity theorem: the
pdf $\mathbb P^{(T,\infty)}$ converges weakly, as $T\to\infty$, towards a
pdf $\bar{\mathbb P}$ such that
$$
\int_\rr \e^{\i \alpha \Delta Q}\d \bar {\mathbb P}(\Delta Q)
=\bar\varphi(\alpha).
$$
We note that while $\mathbb P^{(T,\infty)}$ is, in general, a continuous pdf, 
$\bar{\mathbb P}$ is atomic.

\noindent{\bf Remark.} From Eq.~\eqref{EQ-WForm}, we infer that the FS of the work
done on the joint system $\cS+\cR$ during the process can be obtained by the successive
measurements of $H^{(L)}(0)$ at the epoch $s=0$ and $H^{(L)}(1)$ at the epoch $s=1$. A simple
modification of the calculation of Section~\ref{SEC-Setup} yields the cumulant generating 
function of the work
$$
\chi_{\rm work}^{(T,L)}(\alpha)=-\alpha\Delta F
+S_{\frac\alpha\beta}(\eta_\f^{(L)}|\rho_{1}^{(T,L)}).
$$
Proceeding as before, one shows that
$$
\lim_{T\to\infty}\lim_{L\to\infty}\chi_{\rm work}^{(T,L)}(\alpha)=-\alpha\Delta F,
$$
which is the cumulant generating function of a deterministic quantity.
Thus, the work done on the system does not fluctuate in the adiabatic limit and 
is equal to the increase of the free energy.

\section{Refinement of Landauer's Principle}

We return to our discussion of the Landauer erasure principle. 
Recall that we
consider the case where $\lambda(0)=\lambda(1)=0$, that the initial state  is $\rho_\i=I/d$, and that 
the final state is  $\rho_\f>0$. 
The difference
between the initial and the final entropy of the system is hence 
$\Delta S=\log d-S_\f$. The adiabatic theorem for thermal states implies that the
time-evolved state $\rho^{(T,\infty)}_1$ converges in the adiabatic limit to
the product state $\eta_\f^{(\infty)}=\rho_\f\otimes\rho_\cR^{(\infty)}$, 
realizing the task of transforming $\rho_\i$ into $\rho_\f$ (here, 
$\rho_\cR^{(\infty)}$ denotes the thermal equilibrium state of $\cR$ at inverse 
temperature $\beta$). We now consider the energetic cost of this transformation.

Let $p_{k}$ denote the eigenvalues of $\rho_\f$ and $m_k$ their respective multiplicities.
We can rewrite the cumulant generating function~\eqref{FCSg} as
\beq
\log\int\e^{-\alpha\Delta Q}\d\bar{\mathbb{P}}(\Delta Q)
=\bar{\chi}(\alpha)
=\log\sum_k\frac{m_k}d \e^{(\beta-\alpha)Q_k}, \quad Q_k=\frac1\beta(\log d +\log p_k),
\label{EQ-chibar}
\eeq
which shows that heat is quantized. A heat quanta $Q_k$ is dissipated in the bath 
with probability
$$
\bar{\mathbb{P}}(\Delta Q=Q_k)=p_k m_k=\frac{m_k}d \e^{\beta Q_k}.
$$
Differentiating~\eqref{EQ-chibar} at $\alpha = 0$, we immediately obtain the saturation 
of the Landauer Principle for the expected heat,
$$
\langle\Delta Q\rangle
=-\partial_\alpha\bar{\chi}(\alpha)\big|_{\alpha=0}
={\beta}^{-1}\Delta S.
$$
The expression for higher cumulants reads
\begin{equation}
\label{moments}
\langle\langle\Delta Q^n\rangle\rangle
=(-\partial_\alpha)^n\bar{\chi}(\alpha)\big|_{\alpha=0}
=\beta^{-n}\partial_\gamma^n\log\sum_km_kp_k^\gamma\big|_{\gamma=1},\qquad(n\ge2).
\end{equation}
\begin{figure}
\label{FIG-chi}
\begin{center}
\includegraphics[scale=0.5]{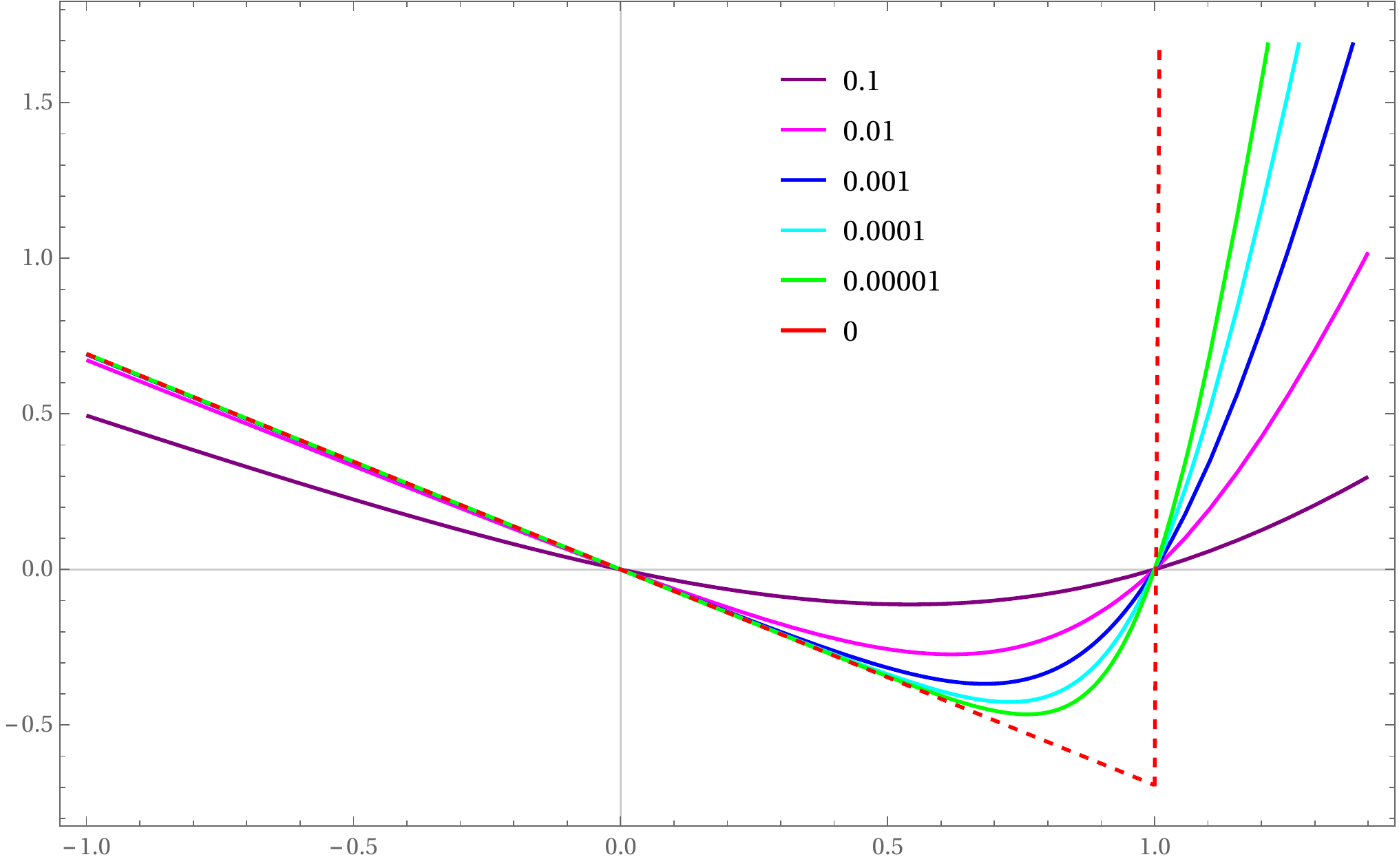}
\end{center}
\caption{The cumulant generating function $\bar{\chi}^{(\epsilon)}$ as a function of 
$\alpha/\beta$ for a qubit ($d=2$) at $\epsilon=10^{-k}$ for $k=1,2,3,4,5$. The 
straight line is the limiting function~\eqref{EQ-LimChi}.}
\end{figure}
Consider now a family of faithful states $\{\rho_\f^{(\epsilon)}\}_{\epsilon\in]0,1/2[}$
such that $\rho_\f^{(\epsilon)}$ approaches a pure state 
$|\psi\rangle\langle\psi|$ as $\epsilon\downarrow0$.
Denote by $\bar{\mathbb{P}}^{(\epsilon)}$ the corresponding heat FS.
Without loss of generality, we can assume that $1-\epsilon$ is an eigenvalue of 
$\rho_\f^{(\epsilon)}$ (with eigenvector $\psi$). Then, this eigenvalue is simple and the rest of the spectrum of 
$\rho_\f^{(\epsilon)}$ is contained in the  interval $]0,\epsilon[$.
Eq.~\eqref{EQ-LimCharF} yields
$$
\lim_{\epsilon\downarrow0}\bar{\varphi}^{(\epsilon)}(\alpha)
=\lim_{\epsilon\downarrow0}d^{\i\frac\alpha\beta}
\tr\left({\rho_\f^{(\epsilon)}}^{1+\i\frac\alpha\beta}\right)=d^{\i\frac\alpha\beta},
$$
which, invoking once again Levy's theorem, implies that $\bar{\mathbb{P}}^{(\epsilon)}$ 
converges weakly to the Dirac mass at $\beta^{-1}\log d$. Thus, in the perfect
erasure 
limit, the heat does not fluctuate either, and takes the value imposed by the 
Landauer 
bound with probability one. 
However, any practical implementation of the erasure process will involve
some errors and the final pure state $\psi$ will only be reached within some precision $\epsilon>0$ (or with some probability $1-\epsilon$).
It is therefore worth paying some attention to the asymptotics $\epsilon\downarrow0$.
In this limit, one easily shows that
$$
\langle\Delta Q\rangle={\beta}^{-1}\log d+\cO(\epsilon\log\epsilon),
$$
while for $n\ge2$, Eq.~\eqref{moments} gives
$$
\langle\langle\Delta Q^n\rangle\rangle=\cO(\epsilon(\log\epsilon)^n).
$$
The presence of powers of $\log\epsilon$ in these formulas is the signature of the 
singularity developed by the cumulant generating function (see Figure~\ref{FIG-chi})
\beq
\lim_{\epsilon\downarrow0}\bar{\chi}^{(\epsilon)}(\alpha)
=\lim_{\epsilon\downarrow0}\log\left(d^{-\frac\alpha\beta}
\tr\left({\rho_\f^{(\epsilon)}}^{1-\frac\alpha\beta}\right)\right)
=\begin{cases}
-\frac{\alpha}{\beta}\log d&\mbox{if }\alpha<\beta,\\
0&\mbox{if }\alpha=\beta,\\
\infty&\mbox{if }\alpha>\beta.
\end{cases}
\label{EQ-LimChi}
\eeq
For small values of $\epsilon$, $d-1$ of the (repeated) eigenvalues of 
$\rho_\f^{(\epsilon)}$ are clustered near zero and the corresponding heat quanta 
become
strongly negative. Accordingly, the system $\cS$ might occasionally absorb large 
amounts of heat $-Q_k^{(\epsilon)}\sim-\beta^{-1}\log\epsilon$. Such heat release 
by the reservoir corresponds to a transition of $\cS$ to an eigenstate
$\phi_k$ of $\rho_\f^{(\epsilon)}$ such that 
$\langle\phi_k|\rho_\f^{(\epsilon)}\phi_k\rangle=\cO(\epsilon)\ll 1$, i.e., to a 
{\sl failure of the erasure process} to reach the pure state $\psi$.
This transition happens at a high energy cost. Thus, it is not surprising that
the fluctuations breaking Landauer's Principle have a total probability 
$\bar{\mathbb{P}}^{(\epsilon)}(\Delta Q\le0)=\epsilon$ which is exponentially small
w.r.t.\;the energy scale $-\log\epsilon$ involved in the process. 
Still we expect these fluctuations  might be relevant in the experimental 
investigation of the Landauer limit for quantum systems.

As an alternative approach to the analysis of perfect erasure,
let us compute the probability distribution of the released heat
conditioned on the fact that a final measurement of the system state confirms
the success of the erasure process. Applying Bayes rule we derive, for finite
$L$ and $T$,
$$
\P^{(T,L)}_{\rm success}(\Delta Q)
=\sum_{e'-e=\Delta Q}
\frac{\tr\left(
P^{(L)}_{e'}U_1^{(T,L)}P^{(L)}_e\eta^{(L)}_iU_1^{(T,L)\ast}P^{(L)}_{e'}
P_{\rm success}
\right)}
{\tr\left(
U_1^{(T,L)}\eta^{(L)}_iU_1^{(T,L)\ast}P_{\rm success}
\right)},
$$
where $P_{\rm success}=|\psi\rangle\langle\psi|\otimes I$ denotes the orthogonal 
projection on the target pure state $\psi$. Since this projection commutes with 
$P_{e'}^{(L)}$, the corresponding cumulant generating function reads
$$
\chi^{(T,L)}_{\rm success}(\alpha)
=\log\left(\frac{\tr\left(\e^{-\alpha H_\cR^{(L)}}U_1^{(T,L)}\e^{\alpha H_\cR^{(L)}}\eta^{(L)}_iU_1^{(T,L)\ast}P_{\rm success}\right)}
{\tr\left(U_1^{(T,L)}\eta^{(L)}_iU_1^{(T,L)\ast}P_{\rm success}\right)}\right).
$$
Proceeding as before, we easily obtain the following expression of the 
conditional cumulant generating function of heat in the thermodynamic and 
adiabatic limits and for the target state $\rho_\f^{(\epsilon)}$,
$$
\chi^{(\epsilon)}_{\rm success}(\alpha)
=\lim_{T\to\infty}\lim_{L\to\infty}\chi^{(T,L)}_{\rm success}(\alpha)
=-\frac{\alpha}{\beta}(\log d+\log(1-\epsilon)).
$$
Thus, conditioning on the success of perfect erasure yields a heat distribution 
which concentrates on 
$$
\beta\Delta Q=\log\left(p_{\rm max}d\right)<\Delta S,
$$
where $p_{\rm max}=1-\epsilon$ denotes the largest eigenvalue of 
$\rho_\f^{(\epsilon)}$.
Again, such a departure from Landauer principle could in principle be checked 
experimentally.

\section{Conclusion}
We have studied the statistics of the heat dissipated in a thermal bath during
the quasi-static realization of a Landauer erasure which transforms a completely
mixed initial state into a faithful final state $\rho_\f$. We have shown that
the dissipated heat is quantized, and interpreted this phenomenon as a fine
version of reversibility for isothermal processes. In the singular limit, when
$\rho_\f$ is close to a pure state $|\psi\rangle\langle\psi|$, the heat 
distribution acquires extreme outliers. With a small but non-zero probability a 
large amount of heat can be absorbed by the system during the erasure process. 
This singularity can be detected in the divergence, Eq.~\eqref{1}, of the moment 
generating function of the heat Full Statistics and corresponds to a failure of
the process to reach the pure state $\psi$. Alternatively, conditioning on the 
success of the perfect erasure process yields a heat distribution which is 
concentrated on a value strictly smaller that Landauer's limit.

We believe this departure could be experimentally detected in a quantum analog
of the experiments confirming Landauer's Principle~\cite{SLKL,PBVN,Ra,TSU,BAP}.
Several interferometry and control protocols to measure the heat Full Statistics
using an ancilla coupled to the joint system $\cS+\cR$ were proposed~\cite{CBK,
DCH, GPoM,  MDCP, RCP}. The proposal of Dorner {\sl et al.}~\cite{DCH} seems to
be the most appropriate for our model since it involves only local interactions
between the ancilla and the reservoir.

{\bf Acknowledgment.} The research of T.B. was partly supported by ANR
project RMTQIT (Grant No. ANR-12-IS01-0001-01) and by 
ANR contract ANR-14-CE25-0003-0.  The research of V.J. was partly supported by NSERC. A part 
of this work has been done during a visit of T.B. and M.F. to McGill University 
partly supported by NSERC. The work of C.-A.P. has been carried out in the framework 
of the Labex Archim\`ede (ANR-11-LABX-0033) and of the A*MIDEX project 
(ANR-11-IDEX-0001-02), funded by the ``Investisements d'Avenir'' French 
Government program managed by the French National Research Agency (ANR).


\end{document}